\documentclass[conference]{IEEEtran}
\IEEEoverridecommandlockouts
\usepackage{cite}
\usepackage{amsmath,amssymb,amsfonts}
\usepackage{algorithmic}
\usepackage{graphicx}
\usepackage{textcomp}
\usepackage[table,xcdraw]{xcolor}
\usepackage{balance}
\usepackage{booktabs}
\usepackage{multirow}
\def\BibTeX{{\rm B\kern-.05em{\sc i\kern-.025em b}\kern-.08em
    T\kern-.1667em\lower.7ex\hbox{E}\kern-.125emX}}
\begin{document}

\title{Dark Experience for Incremental Keyword Spotting
}

\author{\IEEEauthorblockN{Tianyi Peng\IEEEauthorrefmark{1} and Yang Xiao\IEEEauthorrefmark{1}\IEEEauthorrefmark{2}}
\IEEEauthorblockA{\textit{\IEEEauthorrefmark{1}Nanyang Technological University} \\
\textit{\IEEEauthorrefmark{2}Fortemedia Singapore, Singapore}\\
Email: \{tpeng003, yxiao009\}@e.ntu.edu.sg}}
\maketitle

\begin{abstract}
Spoken keyword spotting (KWS) is crucial for identifying keywords within audio inputs and is widely used in applications like Apple Siri and Google Home, particularly on edge devices. Current deep learning-based KWS systems, which are typically trained on a limited set of keywords, can suffer from performance degradation when encountering new domains, a challenge often addressed through few-shot fine-tuning. However, this adaptation frequently leads to catastrophic forgetting, where the model's performance on original data deteriorates. Progressive continual learning (CL) strategies have been proposed to overcome this, but they face limitations such as the need for task-ID information and increased storage, making them less practical for lightweight devices. To address these challenges, we introduce Dark Experience for Keyword Spotting (DE-KWS), a novel CL approach that leverages dark knowledge to distill past experiences throughout the training process. DE-KWS combines rehearsal and distillation, using both ground truth labels and logits stored in a memory buffer to maintain model performance across tasks. Evaluations on the Google Speech Command dataset show that DE-KWS outperforms existing CL baselines in average accuracy without increasing model size, offering an effective solution for resource-constrained edge devices. The scripts are available on GitHub~\footnote{https://github.com/NeverlandCookies/DE-KWS} for future research. 
\end{abstract}


\section{Introduction}
Spoken keyword spotting (KWS)~\cite{mandal2014recent, chandra2015keyword, lopezespejo2021deep,10661143} aims to identify specific keywords within audio input. It is a crucial component in many widely-used applications, such as Apple Siri and Google Home, which are commonly deployed on edge devices~\cite{zhang2018hello}. 
Since the KWS system always powers on, maintaining the performance with the small-footprint model is crucial for real-world applications.  Current KWS systems~\cite{chen2014kws,berg2021kwt,kim2021broadcasted,ng23b_interspeech}, based on deep learning, are typically trained with a limited set of keywords to reduce computation and memory usage. Chen et al.~\cite{chen2014kws} were the first to apply deep neural networks to treat keyword spotting as a classification task. Choi et al.~\cite{choi2019temporal} proposed TC-ResNet, a temporal convolutional neural network for real-time KWS on mobile devices, which reduces computation and improves accuracy compared to 2D convolutions. However, this approach can lead to significant performance degradation when the model encounters unfamiliar keywords from a new domain during real-world use.

To address performance degradation, prior research has employed few-shot fine-tuning~\cite{awasthi2021teaching,2021Few, parnami2022few,pan2009survey} to adapt KWS models to new scenarios using target-domain training data. However, this approach often leads to poor performance on the original source-domain data, a phenomenon known as catastrophic forgetting~\cite{mccloskey1989catastrophic}. This occurs because new knowledge gradually replaces previously learned information.

To solve catastrophic forgetting, continual learning (CL)~\cite{clsurvery1,clsurvey2,clsurvey3} aims to continuously acquire new knowledge while retaining and reusing previously learned information. CL has been widely applied in speech processing~\cite{cdoa,ucil,cl3}, audio classification, and other fields. It can be divided into task-incremental and class-incremental learning~\cite{til}, depending on the presence of a task ID. Recent strategies like progressive continual learning~\cite{pclkws2022} apply task-incremental continual learning to KWS, enabling models to adapt to new data while retaining previous knowledge. However, despite these advancements~\cite {delange2021continual}, two key limitations remain the reliance on task-ID as auxiliary information, which is not always available, and the growing storage requirements as tasks increase. In this paper, we focus on class-incremental learning, which is more practical as it does not rely on task IDs. In continual learning, class-incremental learning (CIL)~\cite{cil1,cil2,cil3} integrates new classes into a model’s learning architecture sequentially. Although it has a higher potential for real-world applications, CIL in KWS remains a less-explored area. Recently, Xiao et al.~\cite{cl2} proposed a diversity-based sampler integrated with the self-distillation approach for future rehearsal, and Yang et al.~\cite{yang2023dual} introduced a replay-based dual-memory multi-modal framework for continual KWS tasks. However, these approaches primarily focus on preserving knowledge from previous tasks and often overlook the richer context within speech clips. Dark experience replay (DER)~\cite{der}, which stores past logits to distill experiences over the entire training trajectory, has been suggested as a method, but its application in the KWS domain remains unexplored. 

In this paper, we present a CIL approach for KWS called Dark Experience for Keyword Spotting (DE-KWS). it relies on dark knowledge~\cite{hinton2014dark, hinton2015distilling} for distilling past experiences, sampled over the entire training trajectory. DE-KWS combines rehearsal and distillation, using both ground truth labels and dark knowledge embedded in the logits. We create a memory buffer that stores samples, labels, and logits throughout the training process. The ground truth labels help the model remember past samples through experience replay, while the logits ensure the model retains its response to previous data. Following prior work's setup, we tested DE-KWS on the Google Speech Command dataset and found that it outperforms other CIL baselines in average accuracy without increasing the network's parameters. Our code is made available
at the GitHub. 

\begin{figure}[t]
  \centering
  \includegraphics[width=\linewidth]{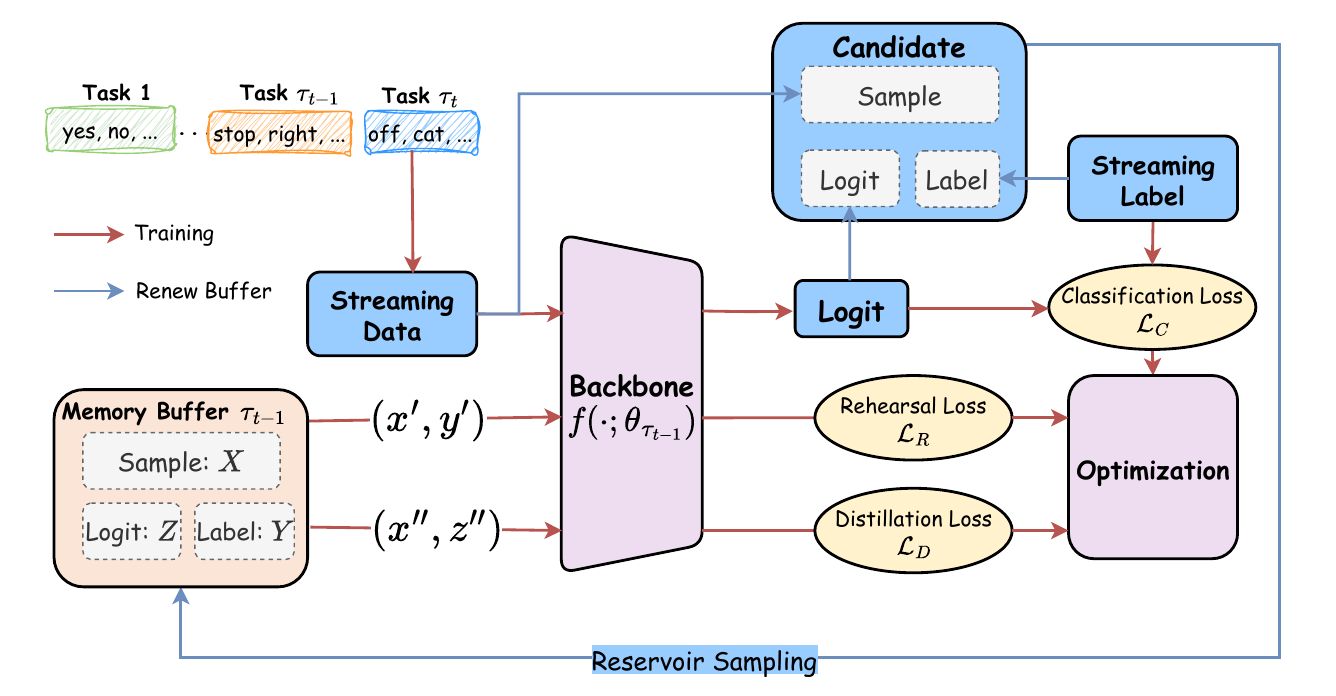}
  \vspace{-8mm}
  \caption{The diagram of the proposed DE-KWS framework for incremental KWS. A buffer $\mathcal{M}_{\tau _{t-1} }$ is deployed to record the memory of the past tasks (i.e., input samples and their corresponding output logits and ground truth labels). The logits obtained from the streaming data via the backbone are used to update the buffer through reservoir sampling. It should be noted that $\tau _{t-1}$ here refers to the status of the buffer and backbone network before the current task $\tau _{t}$}
  \label{fig:workflow}
  \vspace{-5mm}
\end{figure}

\section{Our Method}

\subsection{Problem Formulation}
In this work, we focus on a KWS system that learns different categories of keywords from a sequence of tasks $\left \{ \tau _{1}, \tau _{2},\dots ,\tau _{T} \right \}$. The KWS system aims to recognize all keywords across these tasks, framing it as a CIL problem. For each task $\tau _{t}$, the input audio samples $x$ and their corresponding ground truth labels $ y$ follow the distribution $D_{t}$. The goal of CIL is to train a model $f(x ;\theta )$ that can continually adapt to new data while retaining its memory of old data, ultimately achieving the best overall average accuracy:
\begin{equation}
	\label{eq1}
	\underset{\theta }{\mathrm{argmin}} \sum_{t = 0}^{T}\mathbb{E}_{\left ( x,y   \right )\sim D_{t}  }\left [ \mathcal{L}_{CE}\left ( y, f(x;\theta)  \right ) \right ],
\end{equation}
where $\mathcal{L}_{CE}$ denotes the cross-entropy loss.

Considering data storage costs and time efficiency, full access to past task data and joint training from scratch are both impractical. However, simply fine-tuning on new data can lead to catastrophic forgetting, where the model becomes overly focused on the new data, leading to a significant drop in performance on previously learned tasks.

\subsection{DE-KWS Method}
With these considerations, we introduce the proposed Dark Experience for Keyword Spotting (DE-KWS) method. As shown in Fig.~\ref{fig:workflow}, our model incorporates a memory buffer $\mathcal{M}_{\tau _{t-1} }$ to store experiences from past tasks up to task $\tau _{t-1}$. 

\subsubsection{\textbf{Rehearsal}} Following the replay-based CIL methods, sampled input audio utterances and their corresponding labels are stored in the buffer, and during training, we randomly selected them from the buffer as training data. Instead of mixing the rehearsal data with the current task data to form a unified training dataset, our approach uses them to construct an additional rehearsal loss term to facilitate the model’s retention of memory for past tasks:
\begin{equation}
	\label{eq2}
	\mathcal{L} _{R}=\mathbb{E} _{\left ( x',y' \right )\sim \mathcal{M}_{\tau _{t-1} } } \left [ \mathcal{L}_{CE}\left ( y',f(x' ;\theta_{\tau _{t-1} }  ) \right )  \right ].
\end{equation}
where $\left ( x',y' \right )$ are small batches of data pair sampled from the buffer to compute the rehearsal loss $\mathcal{L}_{R}$.
\subsubsection{\textbf{Distillation}} Apart from rehearsal methods, distillation is also commonly used in incremental learning practices. Distillation is usually applied through self-distillation, where the model’s snapshot from earlier tasks acts as the teacher, and the current model serves as the student. Self-distillation typically involves saving model parameters. The process compares the responses of the past and current models on the data from the current task to determine if the model retains knowledge from previous tasks. However, our approach stores the model’s outputs (i.e., logits), reducing the memory costs of saving past model parameters and reusing the input audio utterances from the rehearsal process. By storing logits instead of full model parameters, our method leverages dark knowledge—implicit information about class relationships embedded in the output logits, which is not captured by ground truth labels.

Building on this approach, we implement a strategy for sampling and using logits continuously throughout the training process, rather than only at the end of one task. This enables DE-KWS to operate without relying on task boundaries, making it more applicable to real-world incremental learning scenarios where task boundaries are often blurry. By sampling logits across the training trajectory, the model captures its progressive evolution, resulting in smoother transitions between tasks and better adaptability. We use reservoir sampling~\cite{vitter1985random} to ensure that each data point has an equal chance of being stored in the buffer, even when the data stream length is unknown. The stored logits and corresponding inputs are randomly sampled to construct a distillation loss term, ensuring the current model’s response remains consistent with previous logits and preserves past task memory.
\begin{equation}
	\label{eq3}
	\mathcal{L}_{D}=\mathbb{E}_{\left ( x'',z'' \right )\sim \mathcal{M}_{\tau _{t-1} } } \left [ \mathcal{L}_{MSE}\left ( z'',h(x'';\theta_{\tau _{t-1} }) \right )    \right ],
\end{equation}
where $\mathcal{L}_{MSE}$ denotes the mean squared error, and $ h(\cdot ;\theta_{\tau _{t-1} })$ denotes the pre-softmax output of the network. The $\left ( x'',z''\right )$ are small batches of data pair sampled from the buffer to compute the distillation loss $\mathcal{L}_{D}$.

\subsubsection{\textbf{Training overview}}
To summarize our approach, the pipeline can be concluded as follows. First, the streaming data of the current task is fed into the model, which predicts logits and computes the loss $\mathcal{L}_{C}$ for the current task. Then, the streaming data, labels, and the corresponding logits are sampled to populate the buffer via reservoir sampling. We sample data pairs from the buffer twice to compute the rehearsal loss $\mathcal{L}_{R}$ and distillation loss $\mathcal{L}_{D}$, respectively. Three loss terms are combined to update the model. The overall objective function of our approach can be written as:
\begin{equation}
	\label{eq4}
	\mathcal{L}_{DE-KWS}=\mathcal{L}_{C} +\alpha \mathcal{L}_{R} +\beta \mathcal{L}_{D},
\end{equation}
where $\alpha $ and $\beta $ are the hyperparameters balancing the loss.

\section{Experiment Setting}
\subsection{Dataset}
We conduct experiments on the \textit{Google Speech Command} dataset v1 (GSC) ~\cite{gsc}, a widely used collection of short audio clips designed for training and evaluating KWS models. The dataset consists of $64,727$ one-second-long utterances of 30 different command words. Following previous work~\cite{cl2,yang2023dual}, we split the dataset into two parts, using $80\%$ for training and $20\%$ for validation, with all audio clips sampled at 16kHz.

\subsection{Backbone Network}
We feed the extracted Mel-frequency cepstrum coefficients (MFCC = 40) into the model. And we adopt the TC-ResNet-8~\cite{choi2019temporal} as the backbone to evaluate our proposed DE-KWS method. TC-ResNet-8 is a lightweight convolutional neural network specifically designed for KWS tasks, optimized for deployment on resource-constrained devices. The network includes 3 residual blocks, which consist of 1D temporal convolutional layers, batch normalization layers and ReLU activation layers, and $ \left \{16,24,32,48  \right \}$ channels for each layer including the first convolutional layer.

\subsection{Reference Baselines}
To demonstrate the performance of our approach, we set nine baselines for comparison. Finetuning refers to training where the model continually adapts to new task data, which often results in overfitting the new data and consequently forgetting the old data. Therefore, we set it as the lower-bound baseline. In contrast, joint training uses data from all tasks for training, making it the upper-bound baseline.
\begin{figure*}[t]
  \centering
  \includegraphics[width=\linewidth]{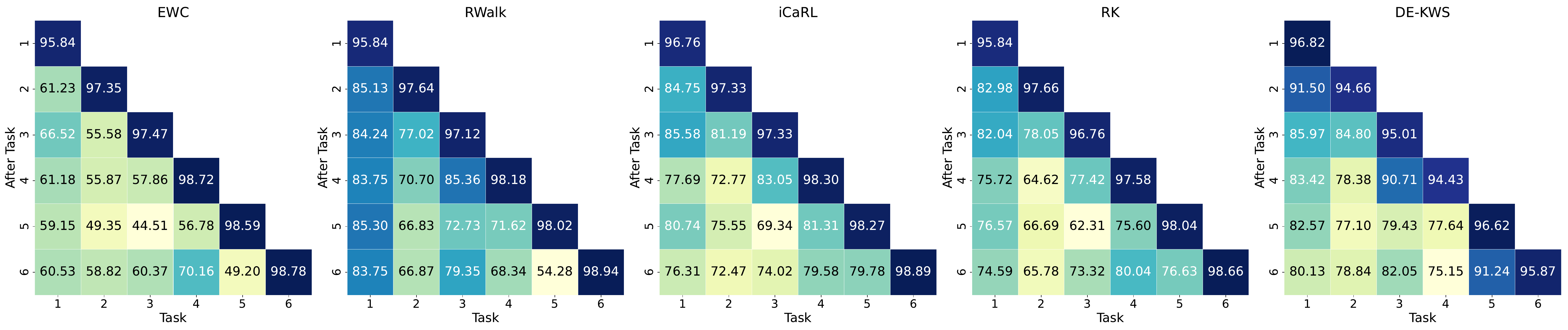 }
  \caption{The task-wise performance comparison of different methods with 500 buffer size.}
  \label{fig:heatmap}
  \vspace{-5mm}
\end{figure*}

We adopt five CIL methods in our incremental KWS task setting. Naive rehearsal (NR)~\cite{hsu2018nr} stores previous data randomly with a replay buffer and uses the stored data for training by an equal amount of new data. The iCaRL~\cite{rebuffi2017icarl} implements its classification functionality with stored samples following the nearest-mean-of-examplars rule and updates the model with an additional knowledge distillation term. EWC~\cite{ewc} mitigates forgetting of previous tasks by adding a regularization term that penalizes significant changes to important model parameters. And the importance of the parameter is determined by the fisher information matrix. RWalk~\cite{rwalk} further improves EWC by combining the fisher information matrix and path integral to measure the importance of the parameter, applying dynamic regularization to parameters based on their significance. BiC~\cite{wu2019bic} introduces an additional bias correction layer to maintain a balance between new and old tasks.

We also select two recent CIL approaches in incremental KWS practice as references. RK~\cite{cl2} proposes a diversity-based sampler, which prevents catastrophic forgetting by ensuring the diversity of each category in the sampled data. DM3~\cite{yang2023dual} framework introduces a dual-memory component to retain the memory of previous tasks and a multi-modal component to further capture the invariant representation.

\subsection{Metrics}
We evaluate the performance in terms of Average Accuracy (ACC), Backward Transfer (BWT)~\cite{lopez2017gradient}, and the number of parameters (Parameters). The ACC metric denotes the average accuracy across all learned tasks. The BWT metric describes the effect of learning new tasks on previous tasks, more specifically, BWT indicates whether the learning of new tasks has a positive or negative impact on the performance of prior tasks. The `Parameter' measures the total parameters of the model in the strategy.

\subsection{Training Detail}
To evaluate the incremental learning ability, we randomly split the data into 6 tasks for the default case. The first task consists of 15 unique keywords, while the remaining keywords are divided into 5 tasks, with 3 keywords per task. We experiment with different buffer sizes to explore their impact on performance. We also compare our method performance with the baselines using different task settings, including 11 tasks (with 10 for the first task and 2 per task for the rest) and 21 tasks (with 10 for the first task and 1 per task for the rest). The model is trained using the Adam optimizer with a learning rate of 0.1. The batch size is set to 128, and the training process is carried out for 50 epochs.

\section{Results}
\begin{table}[!b]
\centering
\vspace{-5mm}
\caption{Comparison of CIL methods for keyword spotting, with Finetune as the lower bound and Joint training as the upper bound, based on accuracy (ACC), backward transfer (BWT), and model parameters across varying buffer sizes.}
\label{tab:baseline}
\resizebox{0.9\columnwidth}{!}{%
\begin{tabular}{@{}c|c|ccc@{}}
\toprule
\textbf{Buffer} & \textbf{Method} & \textbf{ACC$(\uparrow)$} & \textbf{BWT$(\uparrow)$} & \textbf{Parameters} \\ \midrule
\multirow{2}{*}{-}   & Finetune & 26.80          & -0.379          & 64.48K  \\
                     & Joint    & 95.70          & -               & 64.48K  \\ \midrule
-                    & EWC      & 75.23          & -0.105          & 129.96K \\ \midrule
\multirow{6}{*}{200} & NR       & 46.45          & -0.219          & 64.48K  \\
                     & iCaRL    & 78.81          & -0.091          & 75.29K  \\
                     & RWalk    & 80.38          & -0.082          & 129.96K \\
                     & RK       & 77.05          & -0.088          & 129.96K \\
                     & DM3      & 83.35          & -0.058          & 193.44K \\
                     & DE-KWS   & \cellcolor[HTML]{C4D5EB}\textbf{85.13} & \cellcolor[HTML]{C4D5EB}\textbf{-0.048} & 64.48K  \\ \midrule
\multirow{7}{*}{500} & NR       & 55.30          & -0.166          & 64.48K  \\
                     & iCaRL    & 84.72          & -0.057          & 75.29K  \\
                     & RWalk    & 87.53          & -0.047          & 129.96K \\
                     & BiC      & 79.96          & -0.082          & 64.48K  \\
                     & RK       & 84.96          & -0.056          & 129.96K \\
                     & DM3      & 88.10          & -0.043          & 193.44K \\
                     & DE-KWS   & \cellcolor[HTML]{C4D5EB}\textbf{89.24} & \cellcolor[HTML]{C4D5EB}\textbf{-0.034} & 64.48K  \\ \midrule
1000                 & DE-KWS   & 90.38          & -0.029          & 64.48K  \\
1500                 & DE-KWS   & \cellcolor[HTML]{C4D5EB}\textbf{91.17} & \cellcolor[HTML]{C4D5EB}\textbf{-0.026} & 64.48K  \\ \bottomrule
\end{tabular}%
}
\end{table}
\subsection{Comparable Study}
Table~\ref{tab:baseline} compares CIL methods in the context of KWS, where Finetune represents the lower bound and Joint training the upper bound. As expected, none of the methods surpass Joint training, which achieves the highest accuracy (95.70\%) without any forgetting due to its setup. Finetune, the lower bound, has the lowest accuracy (26.80\%) and the worst BWT (-0.379), indicating significant forgetting. As a buffer-free method, EWC achieves moderate performance with 75.23\% accuracy and a BWT of -0.105. For a buffer size of 200, DE-KWS leads with 85.13\% accuracy and a small BWT (-0.048), outperforming other methods. As the buffer size increases, DE-KWS continues to show superior results, reaching 89.24\% accuracy at a buffer of 500 with a BWT of -0.034. At buffer sizes of 1000 and 1500, DE-KWS maintains high performance with 91.17\% accuracy and minimal BWT (-0.026), approaching the upper bound of Joint training. This shows that DE-KWS balances high accuracy and low forgetting.

Compared to other methods, DE-KWS consistently demonstrates superior performance in the initial task (T1), retaining the highest accuracy (80.13\%) even after six tasks. This highlights the strength of DE-KWS in maintaining long-term memory and minimizing forgetting. With sequential tasks, DE-KWS also performs well in the recent tasks (e.g., T5 and T6), showing strong retention of knowledge. We observed that some methods show a decline in performance in mid-term tasks (e.g., RK and RWalk in T2). This suggests a challenge in balancing long-term retention and adaptation to new tasks, typical in class-incremental learning. Not like other models, DE-KWS shows less decline in mid-term tasks. As we mentioned above, using logits as dark experience acts as a regularization method, helping the model retain key features from earlier tasks while adapting to new ones. Sampling logits throughout training captures the model’s evolution, allowing smoother transitions between tasks and improving adaptability. This approach reduces the risk of catastrophic forgetting, especially in mid-term tasks, by stabilizing the model's internal representations as it learns new tasks.

\begin{table}[t]
\centering
\caption{Ablation studies of DE-KWS. ``w/o" means without.}
\vspace{-2mm}
\label{tab:abl-table}
\resizebox{\linewidth}{!}{%
\begin{tabular}{ccccc}
\hline
\textbf{Components}  & \textbf{HyperParameters} & \textbf{ACC$(\uparrow)$} &  \textbf{BWT$(\uparrow)$} \\ \hline 
DE-KWS &\(\alpha=1.0, \beta=0.5\) & 87.95 & -0.039\\
DE-KWS &\(\alpha=0.5, \beta=0.5\) & 88.38 & -0.037\\
 DE-KWS &\(\alpha=0.5, \beta=1.0\) & \cellcolor[HTML]{C4D5EB}\textbf{89.24} & \cellcolor[HTML]{C4D5EB}\textbf{-0.034}\\ \hline
\multicolumn{1}{c}{DE-KWS w/o Rehearsal} & \(\alpha=0.5, \beta=1.0\) & 88.02 & -0.036                         \\  
\multicolumn{1}{c}{DE-KWS w/o Distillation} & \(\alpha=0.5, \beta=1.0\) &     84.84    &  -0.056 \\ \hline
\end{tabular}%
}
\end{table}

\begin{figure}[t]
  \centering
  \begin{minipage}[t]{0.49\linewidth}
  \centering
  \includegraphics[width=\linewidth]{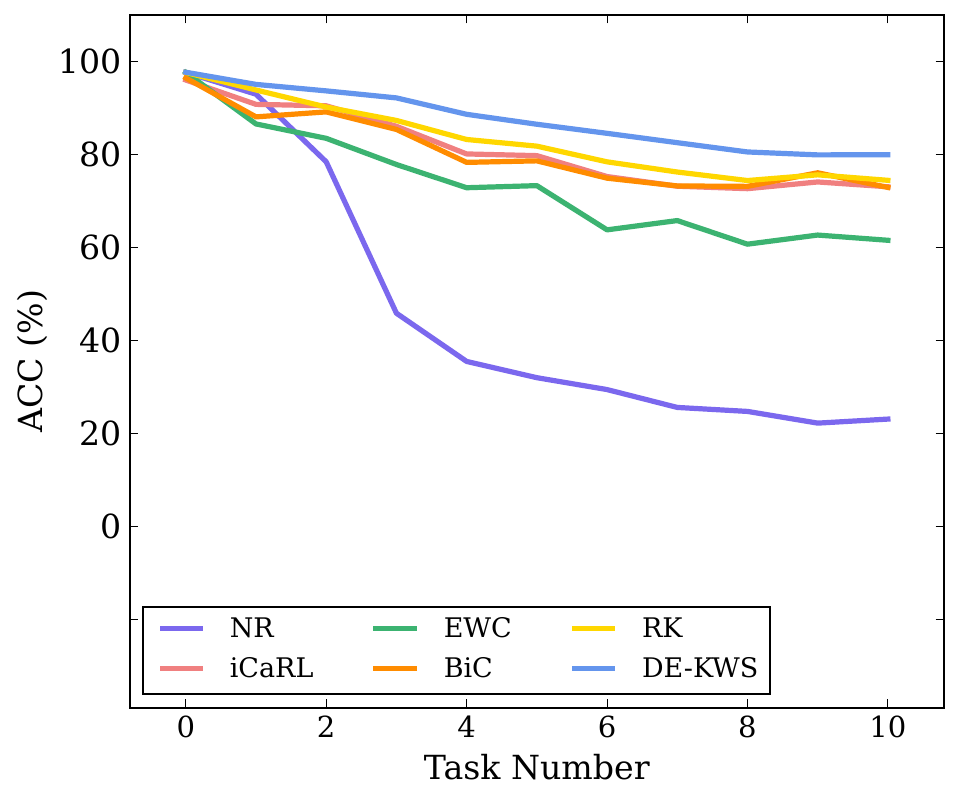}
  \centerline{(a) 10-task scenario}
  \end{minipage}
\hfill
  \begin{minipage}[t]{0.49\linewidth}
  \centering
    \includegraphics[width=\linewidth]{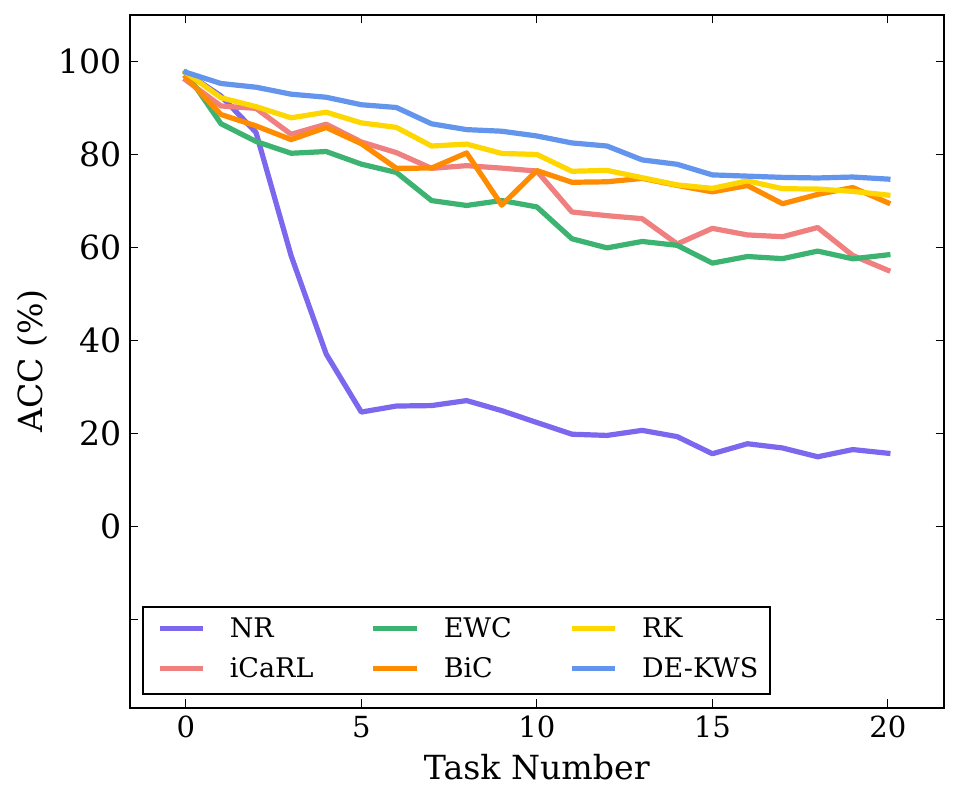}
  \centerline{(b) 20-task scenario}    
  \end{minipage}
  \caption{The ACC (\%) in a comparative study of different task numbers on the proposed DE-KWS approach and other competitive baselines.}
  \label{fig:task}
  \vspace{-5mm}
\end{figure}

\subsection{Ablation Study}
In this ablation study of DE-KWS, different configurations of the hyperparameters \(\alpha\) and \(\beta\) are compared to assess their impact on ACC and BWT. Refer to Table~\ref{tab:abl-table}, with \(\alpha=0.5\) and \(\beta=1.0\) showing the best overall performance in terms of both ACC and BWT.  This suggests that the model achieves the most effective trade-off. The ``without rehearsal" and ``without distillation" experiments demonstrate lower performance, highlighting the importance of both components in DE-KWS for maintaining accuracy and reducing forgetting.

\subsection{Exploration of increasing task numbers}
Fig.~\ref{fig:task} shows the effect of increasing task numbers (10 in (a) and 20 in (b)) on the ACC (\%) for DE-KWS and other competitive baselines. As the task number increases, all methods experience a decline in accuracy, but DE-KWS consistently outperforms the other methods across both task sets. In the 10-task scenario (a), DE-KWS maintains relatively higher accuracy as tasks increase, showing its ability to effectively retain knowledge. In the more challenging 20-task scenario (b), although accuracy declines for all methods, DE-KWS continues to perform better than other methods, particularly in terms of long-term retention. This highlights the robustness of DE-KWS in maintaining accuracy over a larger number of tasks, showcasing its superior handling of catastrophic forgetting in continual learning for KWS.

\section{Conclusion}
In this paper, we introduced Dark Experience for Keyword Spotting (DE-KWS), a novel continual learning approach for keyword spotting that effectively balances knowledge retention and adaptation to new tasks. By leveraging dark knowledge through the use of logits and combining rehearsal with distillation, DE-KWS overcomes the challenges of catastrophic forgetting without increasing model size, making it well-suited for resource-constrained edge devices. Our experiments on the Google Speech Command dataset demonstrate that DE-KWS outperforms existing continual learning baselines in terms of average accuracy. These results highlight the potential of DE-KWS to improve keyword spotting performance in real-world applications while maintaining efficiency.
\clearpage
\balance
\bibliographystyle{IEEEtran}
\bibliography{refs}

\end{document}